\UseRawInputEncoding

\documentclass[twocolumn,floats,floatfix,aps,pra]{revtex4}

\usepackage{amsfonts}
\usepackage{amssymb}
\usepackage{amsmath}
\usepackage{calc}
\usepackage[dvips]{graphicx}
\usepackage{bm}
\usepackage{mathrsfs}
\usepackage{float}

\def\be{\begin{equation}}
\def\ee{\end{equation}}
\def\bea{\begin{eqnarray}}
\def\eea{\end{eqnarray}}
\def\bse{\begin{subequations}}
\def\ese{\end{subequations}}

\def\1{\mathbf{1}}

\begin{document}

\author{Lachezar S. Simeonov}
\affiliation{Department of Physics, Sofia University, James Bourchier 5 blvd, 1164 Sofia, Bulgaria}

\title{Quantum Potential from the Material Derivative of the Osmotic Velocity: A Two-Fluid Madelung Framework}
\date{\today}

\begin{abstract}
We derive the quantum potential directly from the material derivative of the osmotic velocity and formulate a two-fluid model that reproduces the Madelung equations. Interactions between the two fluids are included but remain secondary. The framework is generalized to incorporate electromagnetic fields, yielding a self-consistent description of both the Schr\"{o}dinger and Klein–Gordon equations. Extension to the relativistic Klein–Gordon case demonstrates the model’s flexibility and applicability to spinless relativistic quantum systems. This approach unifies hydrodynamic, kinematic, and electromagnetic perspectives, providing a clear physical interpretation of quantum potentials and forces and offering a versatile platform for modeling complex quantum systems in both non-relativistic and relativistic regimes.
\end{abstract}
\maketitle

\section{Introduction}
The hydrodynamic formulation of quantum mechanics, first proposed by Madelung~\cite{Madelung1927}, revealed that the Schr\"{o}dinger equation can be expressed as a continuity equation coupled to an Euler-type momentum balance containing the now-familiar \emph{quantum potential}. 
This representation suggested that quantum evolution could be viewed as a special form of fluid motion. 
Takabayasi~\cite{Takabayasi1952} extended this framework, interpreting the quantum term in terms of internal stress and vorticity within a ``quantum fluid.''

A complementary stochastic interpretation was introduced by Nelson~\cite{Nelson1966}, who replaced the complex wavefunction by the dynamics of classical particles subject to Brownian fluctuations. 
In this picture, two velocity fields---the \emph{current} and \emph{osmotic} velocities---jointly describe the motion of the ensemble and reproduce the Schr\"{o}dinger equation. 
Related derivations were later developed using stochastic variational principles by Yasue~\cite{Yasue1981} and by Guerra and Morato~\cite{Guerra1983}, establishing the mathematical foundation of stochastic mechanics. 
Although these approaches identified the osmotic velocity as a key dynamical quantity, its explicit convective evolution has remained largely unexplored. 
Subsequent works revisited the role of osmotic and diffusive velocities in quantum hydrodynamics and entropy production~\cite{Heifetz2016,Dedes2022,Chavanis2023}, yet without linking their material derivatives to the emergence of the quantum potential itself.

Beyond stochastic mechanics, the quantum potential has appeared in many alternative derivations. 
In information-theoretic treatments, Frieden~\cite{Frieden1998}, Reginatto~\cite{Reginatto1998}, and Hall and Reginatto~\cite{Hall2002} demonstrated that extremizing Fisher information or enforcing exact-uncertainty relations on a classical ensemble yields the Schr\"{o}dinger equation, with the quantum potential arising as an informational correction. 
Similarly, Caticha’s entropic dynamics framework~\cite{Caticha2019} recovers the same structure from constraints of entropic inference and information geometry. 
In thermodynamic and sub-quantum models, Gr\"{o}ssing~\cite{Grossing2009} and Curcuraci and Ramezani ~\cite{Curcuraci2019} interpreted the quantum term as emerging from hidden thermal or vacuum fluctuations. 
In scale-relativity and fractal-spacetime theories, introduced by Nottale~\cite{Nottale1992} and extended by Chavanis~\cite{Chavanis2023}, it originates from the non-differentiable geometry of particle paths. 
From a kinetic-theory viewpoint, moment closures of the Wigner equation yield the Bohm potential as a higher-order pressure correction~\cite{Cai2012,Jungel2005}. 
Finally, continuum and solid-state analyses interpret it as the divergence of a quantum stress tensor~\cite{Maranganti2010}.

Despite this wide variety of formulations---stochastic, informational, thermodynamic, geometric, or kinetic---the quantum potential is almost always introduced as an \emph{external correction} to classical dynamics. 
In this work, we present a new and purely \emph{kinematic} derivation. 
Starting from the standard diffusion law, the osmotic velocity field emerges naturally. We then show that the quantum potential arises directly from the \emph{material (convective) derivative} of the osmotic velocity along the flow of the current velocity. 
This yields a transparent dynamical interpretation of the quantum force as the inertial response of the osmotic motion, unifying the stochastic and hydrodynamic descriptions within a single fluid framework. 
To our knowledge, this explicit derivation---linking the quantum potential directly to the material derivative of the osmotic velocity---has not been reported previously.

In Section~II, we derive the quantum potential directly from the material derivative of the osmotic velocity and construct a model of two interacting fluids.
One of the fluids obeys the Madelung system, from which the Schr\"{o}dinger equation is recovered.
In Section~III, we extend the model to the relativistic regime and derive the corresponding Madelung equations for the Klein-Gordon equation.
Finally, Section~IV presents conclusions and perspectives.

\section{Material-Derivative Derivation of the Quantum Potential and Two-Fluid Madelung Equations}

\subsection{Quantum Potential from the Material Derivative of the Osmotic Velocity}

Let us consider a fluid with density and velocity fields $\rho$ and $\mathbf{u}$, respectively.
Suppose the fluid is subjected to a diffusion process governed by the standard diffusion law: $\textbf{J}=-D\nabla \rho$, where $\mathbf{J} = \rho \mathbf{u}$ is the current density and $D$ is the diffusion constant. In this case, the fluid velocity corresponds to the familiar osmotic velocity (up to a sign):
\begin{equation}
\textbf{u}=-D\nabla\ln\rho.
\end{equation}
Let us consider the material derivative of the osmotic velocity, i.e., the \emph{osmotic acceleration}:
\begin{equation}
\frac{d\textbf{u}}{dt}=\partial_{t}\textbf{u}+(\textbf{u}\cdot\nabla)\textbf{u},\label{FullDer}
\end{equation} 
The partial derivative is:
\begin{equation}
\partial_{t}\textbf{u}=-D\partial_{t}\nabla\ln\rho=D\frac{\nabla\rho}{\rho^{2}}\partial_{t}\rho-\frac{D}{\rho}\nabla\partial_{t}\rho.
\end{equation}
On the other hand, the continuity equation for the fluid reads $\partial_{t}\rho=-\nabla\cdot \textbf{J}=D\nabla^{2}\rho$. Substituting $\partial_t \rho$ from this equation into the expression above and using Einstein’s summation convention, we obtain, for the $i$th component of $\partial_t \mathbf{u}$,
\begin{equation}
(\partial_{t}\textbf{u})_{i}=\frac{D^{2}}{\rho^{2}}\partial_{k}\partial_{k}\rho\partial_{i}\rho-\frac{D^{2}}{\rho}\partial_{k}\partial_{k}\partial_{i}\rho.\label{TimeDer}
\end{equation}
Next, we compute the $i$th component of the convective acceleration:
\begin{align}
&[(\textbf{u}\cdot\nabla)\textbf{u}]_{i}=\left(-D\frac{\nabla\rho}{\rho}\cdot\nabla\right)\left(-D\frac{\nabla\rho}{\rho}\right)_{i}=\notag\\
&=\frac{D^{2}}{\rho^{3}}\left(\rho\partial_{k}\rho\partial_{k}\partial_{i}\rho-\partial_{k}\rho\partial_{k}\rho\partial_{i}\rho\right).\label{Conv}
\end{align}

Next, substituting Eqs.~\eqref{TimeDer} and \eqref{Conv} into Eq.~\eqref{FullDer}, the full material derivative of the osmotic velocity can be written as:
\begin{equation}
\frac{d\textbf{u}}{dt}=\nabla\left(-2D^{2}\frac{\nabla^{2}\sqrt{\rho}}{\sqrt{\rho}}\right).\label{QuantPot}
\end{equation}
Quite remarkably, the term in the brackets—given a suitable choice of the diffusion constant $D$—coincides with the familiar quantum potential in the Madelung fluid equations.

We will exploit this observation to construct a model of the Madelung fluid equations (and, consequently, the Schr\"{o}dinger equation) in terms of two interacting fluids.

This approach may open a novel and unexplored avenue for fluid-inspired formulations of the Schr\"{o}dinger equation.
Unlike previous stochastic, informational, or geometric approaches, the quantum potential here arises directly from the material derivative of the osmotic velocity, providing a purely kinematic and dynamical interpretation within the fluid framework.

\subsection{Madelung Fluid Equations via a Two-Fluid Interaction}
We shall build on the result that the material derivative of the osmotic velocity naturally gives rise to the quantum potential in the Madelung fluid equations.
In this subsection, we construct a model of a Madelung fluid.

To this end, we consider \textit{two} fluids with density and velocity fields $\rho_a$ and $\mathbf{u}_a$, respectively, where $a = 1,2$.
The two fluids are subjected to a process of mutual diffusion, meaning that $\textbf{J}_{2}=-D\nabla\rho_{1}$ for some diffusion constant $D$ (to be specified below). Then

\begin{equation}
\textbf{u}_{2}=-D\frac{\nabla\rho_{1}}{\rho_{2}}=-D\nabla\ln\rho_{2}+D\frac{\nabla\delta\rho}{\rho_{2}},\label{Diff1}
\end{equation}
where $\delta\rho=\rho_{2}-\rho_{1}$. 

Next, we consider two time scales, as shown in FIG. 1.

\begin{figure}[tb]
\includegraphics[width= 1.0\columnwidth]{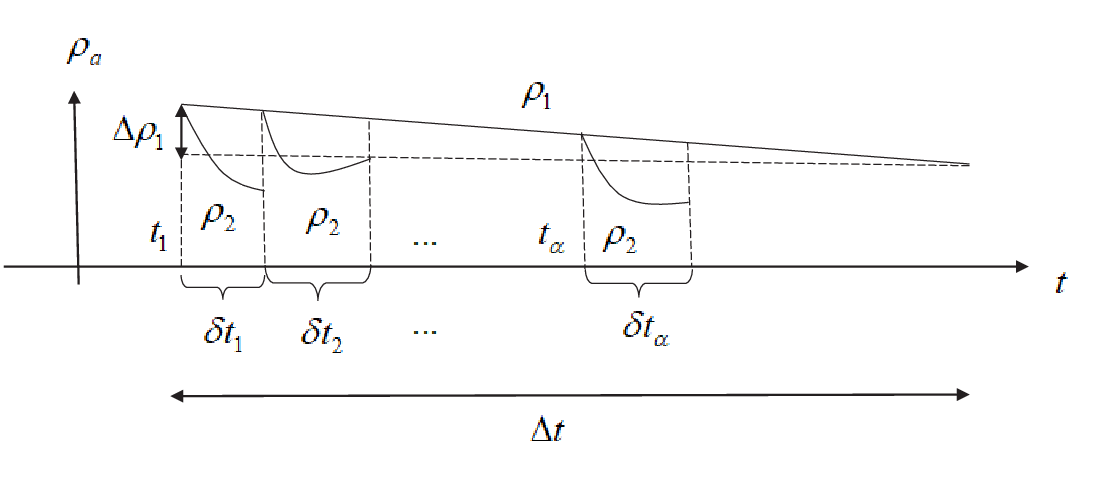}
\caption{Densities $\rho_a$ of the two fluids ($a=1,2$). Two time scales are considered: short intervals $\delta t_\alpha$ ($\alpha=1,2,\dots,N$) and a longer interval $\Delta t$, with $\delta t_\alpha \ll \Delta t$. At the end of each short interval $\delta t_\alpha$, a sudden jump enforces $\rho_2 = \rho_1$, while between jumps $\rho_2$ evolves according to the simple diffusion law. Over the longer time $\Delta t$, the density $\rho_1$ changes by $\Delta \rho_1$, with $|\Delta \rho_1| \ll \rho_1$. The figure is schematic: in reality, $\rho_2$ never deviates significantly from $\rho_1$.}
\label{fig1}
\end{figure}

The first time scale, the interval $\Delta t$, is assumed to be sufficiently small such that during $\Delta t$, the density of the first fluid changes by $\Delta \rho_1$, but $|\Delta \rho_1| \ll \rho_1$.
The second time scales, $\delta t_\alpha$ ($\alpha = 1,2,\dots,N$), are much shorter than $\Delta t$, i.e., $\delta t_\alpha \ll \Delta t$ for each $\alpha$.

We assume that at the end of each short interval $\delta t_\alpha$, a sudden jump enforces the equality of the fluid densities: $\rho_2 = \rho_1$ (cf. Fig.~1).
Between these jumps, fluid 2 evolves according to the simple diffusion law, Eq.~\eqref{Diff1}.
Moreover, within each short interval $\delta t_\alpha$, the density difference $\delta \rho$ in Eq.~\eqref{Diff1} can be neglected, since fluid 2 does not have sufficient time to deviate appreciably from $\rho_1$.

Under these conditions, the velocity of fluid 2 reduces to the standard osmotic velocity, $\textbf{u}_{2}=-D\nabla\ln\rho_{2}$, and its acceleration can be directly obtained from Eq.~\eqref{QuantPot}.
\begin{equation}
\frac{d\textbf{u}_{2}}{dt}=\nabla\left(-2D^{2}\frac{\nabla^{2}\sqrt{\rho_{2}}}{\sqrt{\rho_{2}}}\right);\; t_{\alpha}\leq t\leq t_{\alpha}+\delta t_{\alpha},\; \alpha=1,2,...
\end{equation}

Below, we will require the time average of this osmotic acceleration over the longer interval $\Delta t$, which gives
\begin{align}
&\overline{\frac{d\textbf{u}_{2}}{dt}}=\frac{1}{N}\sum_{\alpha=1}^{N}\frac{d\textbf{u}_{2}(t_{\alpha})}{dt}=\notag\\
&=-2D^{2}\nabla\frac{1}{N}\sum_{\alpha=1}^{N}\frac{\nabla^{2}\sqrt{\rho_{2}(t_{\alpha})}}{\sqrt{\rho_{2}(t_{\alpha})}}=-2D^{2}\nabla\frac{\nabla^{2}\sqrt{\rho_{1}}}{\sqrt{\rho_{1}}}+O(\Delta t).\label{OssmoticAcc}
\end{align}
since $\rho_2(t_\alpha) = \rho_1(t) + O(\Delta t)$. Later, we take the limit $\Delta t \to 0$ and remove $O(\Delta t)$. It then follows that the average force acting on the fluid 2 parcel of volume $\delta V$ is
\begin{equation}
\delta \textbf{F}_{2}=\rho_{2}\delta V \overline{\frac{d\textbf{u}_{2}}{dt}}.
\end{equation}

We now return to fluid 1. By Newton’s third law, the average force per unit mass acting on a fluid 1 parcel of volume $\delta V$ is
\begin{equation}
\frac{\delta \textbf{F}_{1}}{\rho_{1}\delta V}=-\frac{\delta \textbf{F}_{2}}{\rho_{1}\delta V}=-\frac{\rho_{2}}{\rho_{1}}\overline{\frac{d\textbf{u}_{2}}{dt}}=-\overline{\frac{d\textbf{u}_{2}}{dt}}+O(\Delta t).
\end{equation}
Again, we have used $\rho_2 = \rho_1 + O(\Delta t)$.
Consider a particle of mass $m$ subjected to a potential $U$; classically, its acceleration is $-\nabla U / m$. We assume that the entire fluid 1 experiences the same acceleration $-\nabla U / m$ due to the potential $U$. Then, the total acceleration acting on a fluid 1 parcel is
\begin{equation}
\frac{d\textbf{u}_{1}}{dt}=-\frac{1}{m}\nabla U+\frac{\delta \textbf{F}_{1}}{\rho_{1}\delta V}=-\frac{1}{m}\nabla U-\overline{\frac{d\textbf{u}_{2}}{dt}}.\label{Madel0}
\end{equation}
Taking Eq.~\eqref{OssmoticAcc} into account, we finally obtain
\begin{equation}
\frac{d\textbf{u}_{1}}{dt}=-\frac{1}{m}\nabla U+2D^{2}\nabla\frac{\nabla^{2}\sqrt{\rho_{1}}}{\sqrt{\rho_{1}}}.\label{Fluid1}
\end{equation}
If we choose the diffusion constant as,
\begin{equation}
D=\frac{\hbar}{2m},\label{DiffusionCons}
\end{equation}
we then obtain
\begin{equation}
\frac{\partial \textbf{u}_{1}}{\partial t}+(\textbf{u}_{1}\cdot\nabla)\textbf{u}_{1}=-\frac{1}{m}\nabla(U+Q),\label{Madel1}
\end{equation}
where $Q$ is the familiar quantum potential,
\begin{equation}
Q=-\frac{\hbar^{2}}{2m}\frac{\nabla^{2}\sqrt{\rho_{1}}}{\sqrt{\rho_{1}}}.
\end{equation}
Taking into account the continuity equation for fluid 1,
\begin{equation}
\frac{\partial\rho_{1}}{\partial t}+\nabla\cdot(\rho_{1}\textbf{u}_{1})=0,\label{Madel2}
\end{equation}
we finally get the full system of Madelung fluid equations, namely Eqs. \eqref{Madel1} and \eqref{Madel2}.

We make several remarks. First, in Eq.~\eqref{DiffusionCons}, the diffusion constant coincides with that in Nelson’s stochastic mechanics \cite{Nelson1966}. However, whereas Nelson’s model is a hidden-variable theory, our approach is compatible with both hidden-variable and non–hidden-variable interpretations. Furthermore, in Nelson’s framework, the particle undergoes complicated forward and backward Wiener stochastic motions. In contrast, our model is significantly simpler and does not require such stochastic processes. This simplicity represents a major strength of our approach.

Second, we have implicitly taken the time average over the interval $\Delta t$ for fluid 1 as well. Indeed, for any field $X_1$ associated with fluid 1, the time-averaged value of the field is
\begin{equation}
\overline{\frac{dX_{1}}{dt}}=\frac{1}{\Delta t}\int_{t}^{t+\Delta t}\frac{dX_{1}}{dt}dt=\frac{\Delta X_{1}}{\Delta t}.
\end{equation}
We have implicitly assumed this time average throughout our Madelung equations, Eqs.~\eqref{Madel1} and \eqref{Madel2}, which is why the average force $-\overline{d\mathbf{u}_2/dt}$ appears in Eq.~\eqref{Madel0}.
For simplicity, we have suppressed the notation $\Delta X_1 / \Delta t$ and instead written $dX_1 / dt$, since we consider only small $\Delta t$ and ultimately take the limit $\Delta t \to 0$.

Third, the use of \emph{two} fluids is necessary for the following reasons. The convective acceleration $(\mathbf{u}_2 \cdot \nabla) \mathbf{u}_2$ is required to derive the quantum potential $Q$, which clearly indicates that fluid 2 parcels move \emph{independently} of fluid 1. Moreover, the probability distribution of the particle is determined by the density of fluid 1, $\rho_1$, not $\rho_2$; in other words, the particle does not “see” fluid 2. This motivates the necessity of introducing two distinct fluids.

We now return to the derivation of the Schr\"{o}dinger equation. To this end, we consider only potential flow, so that we can write $m \mathbf{u}_1 = \nabla S$. Substituting this into Eq.~\eqref{Madel1}, we obtain the familiar quantum Hamilton–Jacobi equation (QHJE), $\partial_t S + (\nabla S)^2 / 2m + U + Q = 0$, where the integration “constant” (actually a function of time) has been absorbed into $S$.

If we define $R = \sqrt{\rho_1}$, the continuity equation becomes $\partial_t R + (2R)^{-1} \nabla \cdot (R^2 \nabla S / m) = 0$. Multiplying the QHJE by $i R / \hbar$ and adding the continuity equation, then multiplying the result by $\exp(i S / \hbar)$ and substituting $\Psi = R \exp(i S / \hbar)$, we finally obtain the Schr\"{o}dinger equation, $i \hbar \partial_t \Psi = - \hbar^2 \nabla^2 \Psi / 2m + U \Psi$.

The treatment of the two-fluid system in a general electromagnetic field proceeds along the same lines as for a single potential. Specifically, in Eq.~\eqref{Fluid1}, we replace the potential force with the general electromagnetic force acting on fluid 1 parcels. This modifies Eq.~\eqref{Fluid1} accordingly, yielding
\begin{equation}
\frac{\partial\textbf{u}_{1}}{\partial t}+(\textbf{u}_{1}\cdot\nabla)\textbf{u}_{1}=\frac{q}{m}\textbf{E}+\frac{q}{m}\textbf{u}_{1}\times \textbf{B}+2D^{2}\nabla\frac{\nabla^{2}\sqrt{\rho_{1}}}{\sqrt{\rho_{1}}}.
\end{equation}
We adopt the same choice of diffusion constant as before. Let us now introduce the standard electromagnetic potentials $\phi$ and $\mathbf{A}$ through the familiar substitutions $\mathbf{E} = -\nabla \phi - \partial_t \mathbf{A}$ and $\mathbf{B} = \nabla \times \mathbf{A}$, and by writing $\mathbf{u}_1 = (\nabla S - q \mathbf{A}) / m$. Substituting these into the QHJE, we obtain $\partial_t S + (\nabla S - q \mathbf{A})^2 / 2m + q \phi + Q = 0$.

Similarly, the continuity equation becomes $\partial_t R + (2R)^{-1} \nabla \cdot \big(R^2 (\nabla S - q \mathbf{A}) / m\big) = 0$, where $R = \sqrt{\rho_1}$. Following the same procedure as before—multiplying the QHJE by $i R / \hbar$, adding the continuity equation, and substituting $\Psi = R \exp(i S / \hbar)$—we recover the standard Schr\"{o}dinger equation for a general electromagnetic field, $(2m)^{-1} (\hbar \nabla / i - q \mathbf{A})^2 \Psi + q \phi \Psi = i \hbar \partial_t \Psi$, as expected.

Thus, our model can be straightforwardly generalized to include an arbitrary electromagnetic field.

\section{Relativistic Madelung Equations and the Klein–Gordon Equation}
In this section, we present the relativistic extension of our method. We derive the Madelung equations corresponding to the Klein–Gordon equation by employing a relativistic diffusion law.

We consider the flat metric $\eta_{\mu\nu}$ with signature $(+,-,-,-)$ and use natural units, $c = \hbar = 1$. Let us define the current $J^\mu = \rho u^\mu$, where $\rho$ and $u^\mu$ denote the fluid density and four-velocity, respectively. The relativistic diffusion law is then $J^\mu = + D \partial^\mu \rho$. Note the plus sign, which arises from our choice of metric signature. Under these conditions, we have
\begin{equation}
u^{\mu}=D\partial^{\mu}\ln\rho.
\end{equation}

We can now straightforwardly calculate the acceleration of the fluid,
\begin{equation}
u_{\mu}\partial^{\mu}u^{\nu}=D^{2}\partial_{\mu}\ln\rho\partial^{\mu}\partial^{\nu}\ln\rho.
\end{equation}
By interchanging the derivatives, $\partial^\mu \partial^\nu = \partial^\nu \partial^\mu$, and using the identity valid for any four-vector $\chi$, $\chi^\mu \partial^\nu \chi_\mu = \frac{1}{2}\partial^\nu (\chi^\mu \chi_\mu)$, we obtain
\begin{equation}
u^{\mu}\partial_{\mu}u^{\nu}=\frac{1}{2}D^{2}\partial^{\nu}\left(\frac{\partial^{\mu}\rho\partial_{\mu}\rho}{\rho^{2}}\right).
\end{equation}
Substituting $R = \sqrt{\rho}$, we obtain
\begin{align}
u^{\mu}\partial_{\mu}u^{\nu}&=2D^{2}\partial^{\nu}\left(\frac{\partial_{\mu}R\partial^{\mu}R}{R^{2}}\right)\notag\\
&=2D^{2}\left[\frac{\partial_{\mu}(R\partial^{\mu}R)}{R^{2}}-\frac{\partial_{\mu}\partial^{\mu}R}{R}\right]\notag\\
&=2D^{2}\left[\frac{\partial_{\mu}\partial^{\mu}R^{2}}{2R^{2}}-\frac{\partial_{\mu}\partial^{\mu}R}{R}\right].
\end{align}
However, the first term in the brackets cancels. Indeed, from the continuity equation, $\partial_\mu J^\mu = 0$, and using $J^\mu = D\partial^\mu \rho$, it follows that $\partial_\mu \partial^\mu \rho = 0$, i.e., $\partial_\mu \partial^\mu R^2 = 0$. Consequently, the acceleration of the fluid parcels reduces to
\begin{equation}
u^{\mu}\partial_{\mu}u^{\nu}=-\partial^{\nu}\left(2D^{2}\frac{\square R}{R}\right),\label{RelAcc1}
\end{equation}
where $\square = \partial_\mu \partial^\mu$. The term in the brackets is the familiar quantum potential (for an appropriate choice of $D$) corresponding to the Madelung fluid equations for the Klein–Gordon field.

To obtain the actual Klein–Gordon equation, we again consider two fluids with densities $\rho_a$ and four-velocities $u^\mu_a$, $a=1,2$. We also introduce two time scales, as in the non-relativistic case, with corresponding jumps that equalize the densities. Between these jumps, the evolution of fluid 2 is smooth and follows the relativistic law of mutual diffusion,
\begin{equation}
J_{2}^{\mu}=D\partial^{\mu}\rho_{1}.
\end{equation}
which leads to,
\begin{equation}
u_{2}^{\mu}=D\partial^{\mu}\ln\rho_{2}-\frac{D}{\rho_{2}}\partial^{\mu}\delta\rho,
\end{equation}
where, as before, $\delta \rho = \rho_2 - \rho_1$. Between jumps, the density $\rho_2$ does not have sufficient time to deviate significantly from $\rho_1$. In this case, neglecting $\delta \rho$ between jumps, we can apply our result in Eq.~\eqref{RelAcc1}.
\begin{equation}
u_{2}^{\mu}\partial_{\mu}u_{2}^{\nu}=-\partial^{\nu}\left(2D^{2}\frac{\square R_{2}}{R_{2}}\right),
\end{equation}
where $R_a = \sqrt{\rho_a}$, $a=1,2$.
The total energy–momentum tensor for the two fluids is then given by
\begin{equation}
T_{\text{tot}}^{\mu\nu}=m\sum_{a=1,2}\rho_{a}u_{a}^{\mu}u_{a}^{\nu}.
\end{equation}
We have multiplied by the particle mass $m$ because the densities $\rho_a$ are normalized such that their units correspond to inverse volume, rather than mass per unit volume. Let us first consider the case without external fields. In this situation, the conservation law $\partial_\mu T_{\text{tot}}^{\mu\nu} = 0$ leads to
\begin{equation}
\partial_{\mu}\left(\rho_{1}u_{1}^{\mu}u_{1}^{\nu}\right)=-\partial_{\mu}\left(\rho_{2}u_{2}^{\mu}u_{2}^{\nu}\right).
\end{equation}
Using the fact that $\partial_{\mu}\left(\rho_{a}u_{a}^{\mu}\right)=0$, we obtain,
\begin{equation}
\rho_{1}u_{1}^{\mu}\partial_{\mu}u_{1}^{\nu}=-\rho_{2}u_{2}^{\mu}\partial_{\mu}u_{2}^{\nu}
\end{equation}
Dividing by $\rho_1$ and taking the average over the longer time scale $\Delta t$, we obtain
\begin{equation}
u_{1}^{\mu}\partial_{\mu}u_{1}^{\nu}=-\overline{\frac{\rho_{2}}{\rho_{1}}u_{2}^{\mu}\partial_{\mu}u_{2}^{\nu}}=2D^{2}\partial^{\nu}\frac{\square R_{1}}{R_{1}},
\end{equation}
where $R_1 = R_2 + O(\Delta t)$. We have omitted $O(\Delta t)$, and replaced finite differences for fluid 1 with derivatives; this is why no overline appears on the left-hand side. Choosing $D = 1/2m$ (corresponding to $D = \hbar / 2m$ in conventional units, with $\hbar = 1$ here), we finally obtain
\begin{equation}
u_{1}^{\mu}\partial_{\mu}u_{1}^{\nu}=\frac{1}{m}\partial^{\nu}Q,\label{RelMad}
\end{equation}
where $Q$ is the relativistic quantum potential,
\begin{equation}
Q=\frac{1}{2m}\frac{\square R_{1}}{R_{1}}.
\end{equation}
Equation~\eqref{RelMad} represents the relativistic Madelung equation, together with the conservation law,
\begin{equation}
\partial_{\mu}\left(\rho_{1}u_{1}^{\mu}\right)=0,
\end{equation}
we recover the full Madelung system of equations. We now define
\begin{equation}
u_{1}^{\mu}=\frac{1}{m}\partial^{\mu}S.
\end{equation}
We then readily recover the result (temporarily reinstating $c$ and $\hbar$)
\begin{equation}
\partial^{\mu}S\partial_{\mu}S=m^{2}c^{2}+\hbar^{2}\frac{\square R_{1}}{R_{1}},
\end{equation}
The integration constant is chosen as $m^2 c^2$ so that, in the classical limit $\hbar \to 0$, the on-shell condition for the particles is recovered. We now return to natural units, $c = \hbar = 1$. Using the continuity equation and substituting $\psi = R_1 \exp(i S)$, we recover the Klein–Gordon equation,
\begin{equation}
\left(\square +m^{2}\right)\psi=0.
\end{equation} 
The inclusion of interaction with an external electromagnetic field is straightforward. We simply replace $\partial_\mu T_{\text{tot}}^{\mu\nu} = 0$ with 
\begin{equation}
\partial_{\mu}T_{\text{tot}}^{\mu\nu}=eF^{\nu}_{\;\;\lambda}J_{1}^{\lambda},
\end{equation}
where, as usual, $F_{\mu\nu} = \partial_\mu A_\nu - \partial_\nu A_\mu$ denotes the external electromagnetic field, $e$ is the particle’s charge, and $J_1^\mu = \rho_1 u_1^\mu$. The above equation then becomes
\begin{equation}
u_{1}^{\mu}\partial_{\mu}u_{1}^{\nu}=-\overline{\frac{\rho_{2}}{\rho_{1}}u_{2}^{\mu}\partial_{\mu}u_{2}^{\nu}}+\frac{e}{m}F^{\nu}_{\;\;\lambda}u_{1}^{\lambda}=\frac{1}{m}\partial^{\nu}Q+\frac{e}{m}F^{\nu}_{\;\;\lambda}u_{1}^{\lambda}.\label{Acceleration}
\end{equation}
Substituting $m u_1^\mu = \partial^\mu S - e A^\mu$, the acceleration becomes
\begin{align}
u_{1}^{\mu}\partial_{\mu}u_{1\nu}&=\frac{1}{m}\left(\partial^{\mu}S-e A^{\mu}\right)\partial_{\mu}\frac{1}{m}\left(\partial_{\nu}S-e A_{\nu}\right)\notag\\
&=\frac{1}{m^{2}}\left(\partial^{\mu}S-e A^{\mu}\right)\left(\partial_{\nu}\partial_{\mu}S-e\partial_{\mu}A_{\nu}\right)\notag\\
&=\frac{1}{m^{2}}\left(\partial^{\mu}S-e A^{\mu}\right)\left(\partial_{\nu}\partial_{\mu}S-e\partial_{\nu}A_{\mu}-eF_{\mu\nu}\right)\notag\\
&=\frac{1}{m^{2}}\left(\partial^{\mu}S-e A^{\mu}\right)\partial_{\nu}\left(\partial_{\mu}S-e A_{\mu}\right)-\frac{e}{m}u_{1}^{\mu}F_{\mu\nu}\notag\\
&=\frac{1}{2m^{2}}\partial_{\nu}\left[\left(\partial^{\mu}S-e A^{\mu}\right)\left(\partial_{\mu}S-e A_{\mu}\right)\right]+\frac{e}{m}F_{\nu\lambda}u_{1}^{\lambda}
\end{align}
Using this result, Eq.~\eqref{Acceleration} readily reduces to the relativistic QHJE
\begin{equation}
\left(\partial^{\mu}S-eA^{\mu}\right)\left(\partial_{\mu}S-eA_{\mu}\right)=m^{2}+\frac{\square R_{1}}{R_{1}}
\end{equation}
Using the continuity equation and substituting $\psi = R_1 \exp(i S)$, we finally obtain the standard Klein–Gordon equation in an external electromagnetic field,
\begin{equation}
\left(i\partial_{\mu}-eA_{\mu}\right)\left(i\partial^{\mu}-eA^{\mu}\right)\psi=m^{2}\psi
\end{equation}

\section{Conclusion}
In this work, we have developed a fluid-dynamical framework for quantum mechanics based on the convective evolution of the osmotic velocity. By taking the material (convective) derivative of the osmotic velocity, we derive the quantum potential directly, providing a transparent dynamical interpretation of quantum forces within the Madelung hydrodynamic formulation.

Building on this result, we formulated a two-fluid interacting system. Interactions between the fluids arise naturally but remain secondary and do not drive the main dynamics. The model was further generalized to include electromagnetic interactions, yielding a self-consistent description of charged quantum systems. Extending the framework to the relativistic regime produced a Madelung-like formulation of the Klein–Gordon equation, demonstrating the approach’s applicability beyond non-relativistic dynamics.

Overall, this work unifies hydrodynamic, kinematic, and electromagnetic perspectives within a single coherent framework. It provides a practical tool for modeling complex quantum systems and a conceptually transparent interpretation of quantum potentials and forces, opening new avenues for studying interacting quantum fluids, relativistic quantum dynamics, and the interplay between kinematics and potentials in quantum mechanics.


\section{Acknowledgment}
This research did not receive any specific grant from funding agencies in the public, commercial, or not-for-profit sectors.



\end{document}